# The Cloud Adoption Toolkit:
# Supporting Cloud Adoption Decisions in the Enterprise

Ali Khajeh-Hosseini, David Greenwood, James W. Smith, Ian Sommerville
Cloud Computing Co-laboratory, School of Computer Science
University of St Andrews, UK
{akh, dsg22, jws7, ifs}@cs.st-andrews.ac.uk

**Abstract**
Cloud computing promises a radical shift in the provisioning of computing resource within the enterprise. This paper describes the challenges that decision makers face when assessing the feasibility of the adoption of cloud computing in their organisations, and describes our Cloud Adoption Toolkit, which has been developed to support this process. The toolkit provides a framework to support decision makers in identifying their concerns, and matching these concerns to appropriate tools/techniques that can be used to address them. Cost Modeling is the most mature tool in the toolkit, and this paper shows its effectiveness by demonstrating how practitioners can use it to examine the costs of deploying their IT systems on the cloud. The Cost Modeling tool is evaluated using a case study of an organization that is considering the migration of some of its IT systems to the cloud. The case study shows that running systems on the cloud using a traditional 'always on' approach can be less cost effective, and the elastic nature of the cloud has to be used to reduce costs. Therefore, decision makers have to be able to model the variations in resource usage and their systems' deployment options to obtain accurate cost estimates.

## 1 Introduction

Cloud computing is the latest effort in delivering computing resources as a service. Although there are many definitions of cloud computing [e.g. 1, 2], the US National Institute of Standards and Technology (NIST) has published a working definition that has captured the commonly agreed aspects of cloud computing [3]. This definition describes cloud computing as "a model for enabling convenient, on-demand network access to a shared pool of configurable computing resources (e.g., networks, servers, storage, applications, and services) that can be rapidly provisioned and released with minimal management effort or service provider interaction" [3].

Cloud computing represents a shift away from computing as a product that is owned, to computing as a service that is delivered to consumers over the internet from large-scale data centers – or 'clouds'. Cloud computing is currently being exploited by technology start-ups due to its marketed properties of scalability, reliability and cost-effectiveness. Larger enterprises are also beginning to show an interest in cloud computing due to these promised benefits. However, at present much ambiguity and uncertainty exists regarding the actual realization of these promised benefits, as there is currently much hype, particularly around the cost savings of cloud computing which are based on simplistic assumptions. We believe that in the long-term, cloud computing is likely to have a profound effect on the ways software is procured, developed and deployed, similar to the effect of moving from mainframes to PCs.

This paper's original contributions are to:
1. Highlight the challenges of cloud adoption in the enterprise and show that decisions on migrating IT services to the cloud should not simply be driven by cost considerations but should also take a range of socio-technical factors into account.
2. Propose a *Cloud Adoption Toolkit*, which provides a collection of tools that support decision making during the adoption of cloud computing in an enterprise.

Our toolkit is based on a framework to organize thinking about decision makers' concerns and match these to tools that address these concerns, where each tool enables decision makers to focus on and model different attributes of their organizations or IT systems. These models can then be used to reason about and investigate cloud adoption decisions. For example, by modeling hardware infrastructure and applications, it becomes possible to estimate the costs of running that system in the cloud, and hence decide whether deploying that system infrastructure in the cloud would be cost effective. Furthermore, by identifying the impacts of a proposed system to peoples' work activities, its practical and socio-political feasibility can be determined. For instance, a system may be cost effective yet socio-politically infeasible if it potentially decreases job satisfaction and undermines existing power bases or organizational values [4].

This paper starts by highlighting the challenges of cloud adoption in the enterprise (Section 2), and introducing the Cloud Adoption Toolkit (Section 3). We then focus on infrastructure cost modeling and whether it makes financial sense to use the cloud. A case study is used to evaluate the Cost Modeling tool, and show that the cost saving estimates that are often cited by cloud providers cannot be generalized across all IT systems as they depend on the specific resource usage and the deployment options being used by a system (Section 4). The paper concludes by comparing the results of this case study with another case study that was carried out recently, and describing our future work (Section 5).

## 2 Challenges of Cloud Adoption

### 2.1 Problem Overview

The adoption of cloud computing is an emerging challenge that enterprises face in the near-term as the economics of cloud computing become more attractive over time due to economies of scale and competition amongst providers. Companies such as Amazon, Google and Microsoft are investing vast sums in building their public clouds and they seem to be leading the way in the technological innovation of clouds by releasing frequent updates and new features for their services. For example, a quick look at Amazon Web Services' (AWS) news archive[1] shows that they rolled-out over 10 new and technologically impressive features to their cloud offerings in 2009. AWS also released a Security[2] and an Economic[3] center on their website, which shows that there is user demand for advice about the implications of using cloud computing. There is an opportunity for the research community to address this demand by providing independent and impartial advice, tools and techniques to enterprise users who are interested in cloud adoption.

---

[1] http://aws.amazon.com/about-aws/whats-new/2009/
[2] http://aws.amazon.com/security/
[3] http://aws.amazon.com/economics/

The adoption of cloud computing in enterprise environments is non-trivial. Understanding the organizational benefits and drawbacks is far from straightforward because the suitability of the cloud for many classes of systems is unknown or an open-research challenge; cost calculations are complicated due to the number of variables comprising inputs to the utility billing model of cloud computing; the adoption of cloud computing results in a considerable amount of organizational change that will affect peoples' work in significant ways and corporate governance issues regarding the use of cloud computing are not well understood.

### 2.1.1 Cost Calculations

Understanding the operational costs of public clouds is complicated because the cloud's utility billing model is a shift away from capital to operational budgeting, and utility billing has a certain degree of uncertainty that makes it non-trivial to estimate, compared to hardware acquisition. The uncertainty relates to: i) the actual resources consumed by a system, which are determined by its load; ii) the deployment option used by a system, which can affect its costs as resources such as bandwidth are more expensive between clouds compared to bandwidth within clouds; iii) the cloud service provider's pricing scheme, which can change at any time. The consequence is that decision makers are faced with much uncertainty regarding the best provider and whether cloud adoption is more cost effective than other more traditional forms of IT provisioning such as co-location.

Understanding the operational costs of private cloud is also becoming increasingly difficult due to the increasing significance of energy costs and carbon emissions. Concerns for rising energy costs that may be exacerbated by government led carbon taxes [5]. For example, it is predicted that, by 2015, the operational costs of IT infrastructure could exceed its initial capital purchase costs over a 5-year lifecycle [6, 7]. This research challenge is particularly important to cloud computing as its centralized resource-sharing paradigm could be leveraged to optimize energy efficiency.

### 2.1.2 Organizational Change

Understanding the significance and the extent of the organizational changes associated with cloud adoption is a difficult challenge. We argue that enterprises need to understand the breadth of changes and the effort required to make these changes in order to understand their benefits, risks and effects. The success of cloud adoption "is as much dependent on the maturity of organizational and cultural (including legislative) processes as the technology, per se" [8]. The process is likely to be prolonged and some predict that it could take between 10 to 15 years before the typical enterprise makes this shift [9].

A broad number of changes will arise throughout an organisation:
- Accounting will change because hardware and network infrastructure is not procured upfront; it will be consumed as a service and paid for just like a utility.
- Security will change because virtualization introduces new vulnerabilities [10], and there could be conflicts between customers and cloud providers who are both attempting to harden their security procedures [11].
- Compliance will change because the geographic location of data will not be exactly known in the cloud; this has long-term implications for enterprises concerned with data privacy [12, 13, 14].
- Project management will change because the authority of the IT department is going to be eroded by cloud computing. Cloud computing is increasingly turning "users into choosers" [15], and project managers can replace the services provided by the IT department with services offered in the cloud. This is already starting to happen, for example in BP, where a group bypassed the company's IT department by using Amazon Web Services to host a new customer facing website[4].

- System support will change because administrators will no longer have complete control of a system's infrastructure anymore. Their work could increasingly involve contacting cloud providers and waiting for them to look into system problems. Such a scenario was recently reported by Jesper[5] whose application, which was running on Amazon EC2, came under a denial of service attack and had to wait over 16 hours before the problem was fixed.
- Finally, what about the work of end users? The cloud might help collaborative work but what can users do when the cloud goes down? They cannot tell Google or Amazon to prioritize their problem as they could before with their IT department.

The organizational changes will not be straightforward and will require a great deal of management effort due to the highly interconnected nature of legacy infrastructures, the political nature of IT facilitated organizational transformation, and the difficulties of aligning technical systems and organizations.

Large enterprises inevitably have highly interconnected infrastructures comprising a large number of computing systems that have been developed over a long period of time. These depend on different technologies, have different 'owners' within the enterprise and have complex dependencies both between the systems themselves, the data that they process, the middleware used and the platforms on which they run. Business processes have evolved to make use of the portfolio of systems available and these often rely on specific system features. Normally, there is no individual or group within the enterprise who knows about all of the systems that are in use, and dependencies are often discovered by accident when something simply stops working after a change has been made. For international companies, different jurisdictions mean that the same system in different countries has to be used and supported in different ways.

Furthermore, IT provision is profoundly affected by political considerations [16, 17]. Senior management in the enterprise may set IT policies but these are left to individual parts of the enterprise to enact in their own way [18]. Managers naturally tend to adopt strategies that benefit their part of the company and vice versa. Employees resist changes that originate from other parts of the organization [19]. At the inter-group level, the tension between central IT provision and end users has been constant since the 1960s with complaints from users that central services are unwilling or unable to respond quickly to changing user requirements.

The alignment of IT in large organizations is inherently complex, for the reasons mentioned above. For cloud computing to deliver real value, it must be aligned to the enterprise rather than simply be a platform for simple tasks, such as application testing or running product demos. Therefore the issues around migrating application systems to the cloud and satisfying the requirements of key system stakeholders have to be explored. These stakeholders include technical, project, operation and financial managers as well as the engineers who are going to be developing and supporting the systems. Cloud computing is not simply a technological improvement of data centers but a fundamental change in how IT is provisioned and used [20]. Therefore, the adoption of cloud computing will change the work of various system stakeholders in the enterprise, and this will require considerable effort.

---

[4] http://www.trustedcloudservices.com/Individual-Case-Studies/bp-fuels-cloud-computing-interest
[5] http://blog.bitbucket.org/2009/10/04/on-our-extended-downtime-amazon-and-whats-coming/

Cloud adoption decisions are challenging because of a range of practical and socio-political reasons. It is unlikely that all organizations will completely outsource their back-end computing requirements to a cloud service provider. Rather, they will establish heterogeneous computing environments based on dedicated servers, organizational clouds and possibly more than one public cloud provider. How their application portfolio is distributed across this environment depends not only on technical issues but also on socio-technical factors (e.g., concerns about costs, confidentiality, and control), the impact on work practices and constraints derived from existing business models. Therefore, the challenges that need to be addressed are: i) to provide accurate information on costs of cloud adoption; ii) to support risk management; and iii) to ensure that decision makers can make informed trade-offs between the benefits and risks.

## 2.2 Related Work

Academics are beginning to show interest in the challenges of cloud adoption in the enterprise. A recent review of the academic research in cloud computing revealed that there are currently no mature techniques or toolkits available to support decision making during the adoption of cloud computing in the enterprise [21, 22]. In industry, [23] and [24] are examples of typical offerings from IT consultancies that attempt to fill this gap. Such approaches have two problems: they are based on closed proprietary tools that are not widely available; and they are often accompanied by expensive consultancy periods. In contrast, we argue that given the Cloud Adoption Toolkit, enterprises can assess the feasibility of using cloud computing in their organizations quickly and cheaply without outside consultants. However, the toolkit can also be used by decision makers to verify the claims made by IT consultancies and cloud service providers.

The cloud computing literature has so far examined the costs of using the cloud via individual case studies [4, 25, 26, 27], and researchers such as Walker [28, 29] have laid down some of the theoretical foundations for cost modeling. In addition, Buyya's CLOUDS Lab has developed CloudSim [30], which is a useful toolkit for the modeling and simulation of cloud computing environments. As evident by the use-cases mentioned in [30], CloudSim is more suited to developers who are concerned about the performance of their applications, and cloud providers such as HP who are interested in modeling the properties and resource utilization of data centers. In contrast, the Cloud Adoption Toolkit and specifically the Cost Modeling tool are targeted at decision makers in the enterprise who are interested in deploying medium or large-scale IT systems on the cloud. The Cost Modeling tool fills a gap in the existing research by enabling decision makers to model the deployment of complex IT systems on various clouds, including their deployment options and usage patterns as well as different cloud providers' pricing schemes and any future price changes.

The literature has not so far examined the organizational change issues regarding cloud adoption to a great extent. We recently performed a feasibility analysis of a proposed cloud-based IT system at an SME in the oil and gas industry [4]. We found that despite the promised financial benefits, opportunities to remove tedious work from IT staff and the potential to enter new marketplaces, almost all of the stakeholder groups were neutral or reluctant to support a move to the cloud due to concerns regarding its impact on their work, increased risk of dependence upon third parties and its implications for customer service and support. Therefore, from an enterprise perspective, costs are important but so too are customer relationships, public image, flexibility, business continuity and compliance.

# 3 The Cloud Adoption Toolkit

The Cloud Adoption Toolkit comprises a conceptual framework for organizing decision makers' concerns and a mechanism to incorporate supporting tools for each of these concerns. Decision makers can use any tools/techniques that they wish; however, we are developing and propose to incorporate five tools/techniques that we believe to be potentially useful: Technology Suitability Analysis; Energy Consumption Analysis; Stakeholder Impact Analysis; Responsibility Modeling and Cost Modeling.

## 3.1 Conceptual Framework

The purpose of the conceptual framework for cloud decision making is to organize decision makers' thinking about the concerns that they and other stakeholders have, and the tools that can be used to explore these concerns. It is important that decision makers view the proposed cloud adoption project from multiple stakeholders' perspectives in order to learn from a diverse range of stakeholder concerns and receive a broad range of feedback from the organizational environment. Figure 1 provides an overview of the Cloud Adoption Toolkit and how it can be used.

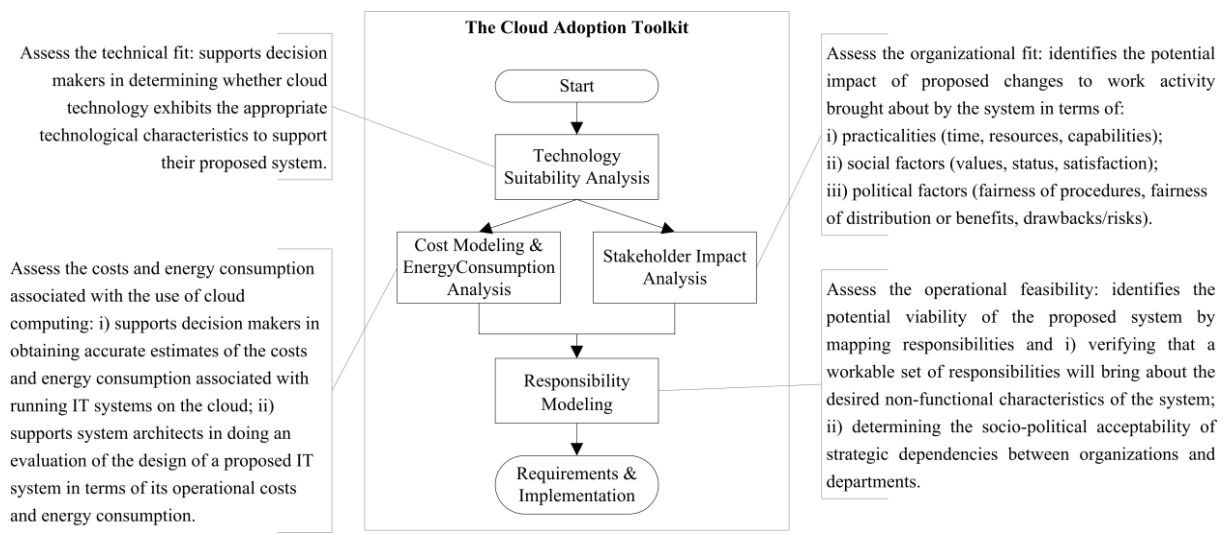

Figure 1: Cloud adoption conceptual framework

Decision makers may start with a Technology Suitability Analysis, and if the cloud is found to be suitable for their system, they then proceed by investigating either the costs of running the system on public clouds, or the energy consumption (and hence energy costs) of running the system on private clouds. At the same time, Stakeholder Impact Analysis can be performed to assess the impacts of using cloud computing on the work of stakeholders in the enterprise. If these analyses show that running the system on the cloud is a viable option, then Responsibility Modeling can be performed to identify and analyze the risks associated with the operation of the system on the cloud, where different cloud providers may be responsible for different aspects of the system.

The process illustrated in Figure 1 is an idealized presentation and, in practice, we expect tool users to make use of the components of our toolkit that they feel are most appropriate. We do not enforce any particular process of use.

Currently, our tool for infrastructure Cost Modeling is the most mature and some support for Technology Suitability Analysis and Stakeholder Impact Analysis is available. Work on Energy Modeling is still at an early stage and we have not yet included a Responsibility Modeling tool. However, Responsibility Modeling is based on an existing notation, developed in a separate project [31, 32], and we are confident that incorporating this into our conceptual framework is a relatively straightforward task.

## 3.2 Technology Suitability Analysis

The purpose of Technology Suitability Analysis is to support decision makers in determining whether cloud computing is the right technology to support their proposed system. Understanding the characteristics of cloud computing is important as it has the potential to exhibit radically different properties to those of traditional enterprise data centers. This is mainly due to the cloud's highly scalable nature, physical resource sharing between virtual machines, potential issues to do with communication over the internet and insufficient guarantees regarding the up-time and reliability of processing and data storage services. For example, typical IaaS offerings make no reassuring guarantees about server uptime or network performance. This has important implications for the viability of certain classes of software architectures and business-critical systems.

Technology Suitability Analysis comprises a simple checklist of questions to provide a rapid assessment of the potent suitability of a particular cloud service for a specific enterprise IT system. The current version of the checklist, shown in Table 1, analyses eight characteristics and quickly provides an indication of the cloud's suitability for a proposed IT system. The outcome of the analysis is a recommendation of whether or not to proceed with further analysis.

**Table 1: Technological Suitability Analysis**

| Desired Technology Characteristic | Questions |
|---|---|
| 1. Elasticity | - Does your software architecture support scaling out?<br>- If not, will scaling up to a bigger server suffice? |
| 2. Communications | - Is the bandwidth within the cloud and between the cloud and other systems sufficient for your application?<br>- Is latency of data transfer to the cloud acceptable? |
| 3. Processing | - Is the CPU power of instances appropriate for your application at the expected operating load?<br>- Do instances have enough memory for the application? |
| 4. Access to hardware / bespoke hardware | - Does your cloud provider provide the required access to hardware components or bespoke hardware? |
| 5. Availability / dependability | - Does your cloud provider provide an appropriate SLA?<br>- Are you able to create the appropriate availability by mixing geographical locations or service providers? |
| 6. Security requirements | - Does your cloud service provider meet your security requirements? (e.g. do they support multi-factor authentication or encrypted data transfer) |
| 7. Data confidentiality and privacy | - Does your cloud provider provide sufficient data confidentiality and privacy guarantees? |
| 8. Regulatory requirements | - Does your cloud provider comply with the required regulatory requirements of your organisation? |

## 3.3 Cost Modeling

The Cost Modeling tool supports the modeling of the costs of running a server infrastructure on the cloud. It therefore supports cloud adoption decisions in two ways:

1. It helps decision makers to obtain accurate cost estimates of running IT systems on the cloud. The tool helps decision makers investigate the costs of migrating an existing IT system or deploying a new IT system on the cloud, the costs of migrating an IT system from one cloud to another, or even future costs based upon predictions of future workload and the provider's pricing scheme.
2. It supports system architects in evaluating the design of a proposed IT system with respect to its operational costs and allows them to compare the costs of different options.

The tool does not currently attempt to take into account the costs of software changes that may be required to gain maximum benefit from using external cloud services. Rather, we focus on infrastructure migration where we assume that existing software can run unchanged on a remote rather than a local server.

The elastic nature of the cloud means that decision makers need a tool to support them in examining the costs of their specific systems as each deployment scenario has different resource usage patterns. The variations in resource usage and a system's deployment options need to be modeled to enable decision makers to consider actual costs rather than the cost saving estimates that are often cited in the press or by cloud providers. As we illustrate later, our tool includes a simple language to describe variable usage patterns for computing resources.

Cost Modeling is based on and extends the capabilities of UML deployment diagrams [33], which enable a system's deployment to be modeled. In its essence, a UML deployment diagram enables users to model the deployment of software artifacts onto hardware nodes. The Cost Modeling tool enables users to model a system's software applications and how they could be deployed on cloud, traditional or hybrid infrastructures. The model is then processed to give users an accurate estimate of the operational costs of their system. The models can take into account future resource demands therefore enabling for situations where traditional infrastructure may not initially be cost effective, yet will become cost effective with future workload increases.

## 3.4 Energy Consumption Analysis

The purpose of Energy Consumption Analysis is to support decision makers in determining the optimum energy consumption of their own private cloud infrastructure. This is important, as there are financial trade-offs to be made between energy efficiency and performance. This tool is currently under development and investigations into this area are ongoing. The inputs of the analysis are the performance per-unit-energy characteristics of physical machines at different levels and types of utilization (e.g. I/O intensive vs CPU intensive) and performance requirements of virtual machine instances. The outputs of the analysis are recommendations for load skewing [34] to ensure that the lowest possible energy is used to meet the virtual machines performance requirements. Additional research is being conducted to understand what kinds of software architectures minimize energy consumption.

## 3.5 Stakeholder Impact Analysis

The purpose of Stakeholder Impact Analysis is to support decision makers in determining the socio-political viability, or benefits and risks, of a proposed IT system. This is important as cloud adoption projects are not merely technological upgrades but involve the reconfiguration of working practices and technologies to take full advantage of the benefits offered by the technology [20]. The socio-political benefits and risks associated with a proposed IT system are determined by identifying the impact of changes to stakeholders' work activities in terms of their practicalities (time, resources, and capabilities), social factors (interests, values, status, and satisfaction) and political factors (their perception of the fairness of decision making procedures and the distribution of benefits, drawbacks and risks). This information enables decision makers to make a judgment about the risk that specific stakeholders will have unsupportive attitudes towards the proposed system and therefore indicates the overall socio-political feasibility of the system. The approach involves:

1. Identifying key stakeholders;
2. Identifying changes in what tasks they would be required to perform and how they were to perform them;
3. Identifying what the likely consequences of the changes are, with regards to stakeholders time, resources, capabilities, values, status and satisfaction;
4. Analyzing these changes within the wider context of relational factors such as tense relationships between individuals or groups to which stakeholders belong;
5. Determining whether the stakeholder will perceive the change as unjust (either procedurally or distributively) based upon changes and their relational context.

We have used Stakeholder Impact Analysis to support decision making in a case study of an oil and gas services company who were interested in exploring the feasibility of migrating an enterprise IT system from an in-house data centre to Amazon EC2 [4]. The analysis revealed that the proposed cloud migration would have many implications for the organization including non-technical areas such as the finance and marketing departments. Overall, a positive net benefit was perceived from the perspectives of the business development functions of the enterprise and the more junior levels of the IT support functions. A zero net benefit was perceived by the project management and support management functions of the enterprise and a negative net benefit was perceived by the technical manager and the support engineer functions of the enterprise. The analysis identified numerous potential benefits and risks associated with the migration. Most notably, opportunities for improved cash flow management, opportunities to offer new products/services, and removal of tedious work were identified as benefits. In contrast, the following notable risks were also elicited: the deterioration of customer care and service quality; increased dependency on external 3rd parties; and departmental down-sizing.

### 3.6 Responsibility Modeling

The purpose of responsibility modeling is to support decision makers in determining the operational viability of a proposed IT system. Responsibility modeling also helps decision makers in identifying and analyzing risks associated with the operation of complex IT systems [31]. Responsibility modeling is particularly important for systems deployed on the cloud as the responsibilities for constructing, operating, maintaining, and managing the system can be divided across multiple organisations, departments and cloud service providers. Thus, identifying and managing the risks associated with the discharge of responsibilities is important to the operational viability of the IT system. Our approach to responsibility modeling is based on an established notation that we have developed and used to model responsibilities in multi-organizational socio-technical systems [31, 32].

The viability of a system is determined by: i) identifying the set of responsibilities that must be discharged for the system to operate according to a set of non-functional requirements; ii) who is responsible for what; iii) whether the configuration of responsibilities is likely to meet non-functional requirements of the system; and iv) determining the practical, social and political viability of the discharge of responsibilities so that the system exhibits appropriate non-functional characteristics e.g. up-time, responsiveness, resilience, maintainability and recoverability.

## 4   Evaluation

Our evaluation of the toolkit is based on case studies, notably a case study of the migration of system infrastructure in an oil and gas services company [4], and in an assessment of the feasibility of migrating part of the IT infrastructure in a higher education institution to the cloud. We focus on this latter example here as it demonstrates the full feature-set of the Cost Modeling tool (the most mature component of the toolkit).

### 4.1   Case Study Overview

The School of Computer Science at the University of St Andrews has around 60 members of staff and 340 undergraduate and postgraduate students. The school provides a number of computing services to its staff and students including:
- Common services such as email, calendar, blog, and web hosting for student projects.
- Storage services such as home directories, backups, and storage of teaching materials.
- Network services such as DNS, VPN, wireless internet and user authentication.

The school has 5 full-time system administrators that maintain its relatively complex IT infrastructure. Some of these systems are interconnected and interact with wider university systems, such as those provided by the university registry and admissions departments. Therefore, the school can be likened to a medium-sized enterprise whose individual systems have evolved over the years to form a mesh of interconnected systems that serve its employees and customers (i.e. the students). The school's computing services are currently deployed on 28 application servers and 5 storage servers in an in-house machine room. There are around 200 desktop machines in the school's computer labs. Some of the school servers are 4 years old and the school is considering upgrading these servers in the near future. The school is considering 3 options:
1. Purchasing new servers to replace the existing servers.
2. Leasing the equivalent amount of resources from the cloud, and migrating its systems but maintaining their existing setup to keep things simple.
3. Leasing resources from the cloud and migrating its systems to the cloud but changing the infrastructure architecture to take advantage of the elasticity of the cloud and so reduce costs.

The Cloud Adoption Toolkit was used to support the school in investigating the feasibility of migrating some of its computing services to the cloud. A review of the school's computing services was carried out to find out which services would be suitable for migration. Technology Suitability Analysis was used as part of this review and the following services were selected as possible candidates for the migration:
- Archive: this service is used by all of the school's storage services and has 560GB of data at the moment.

- StaffRes: this service enables staff to store and manage teaching materials that are used for taught courses.
- StudRes: this service provides read-only access to a subset of the StaffRes files for students to access. StaffRes and StudRes have bursty usage patterns at the beginning and end of the academic year but are not frequently used during the rest of the year.
- Website: the school is thinking of re-building its website as it is outdated. The site sometimes suffers from slow loading times that could be caused by the university network being over utilized.
- WebDev: this service is used for testing the website when it undergoes major updates, but can also be used as a backup if the main web server fails. The website rarely receives major updates; therefore, this service has a small usage.
- WebApps: this service includes blogs, public wikis, and software downloads. These applications are deployed on virtual apache hosts within one of the school servers as they have a very small usage.
- Home directories mirror: this service mirrors the home directories service that provides network storage for all school members and applications. The actual home directories service was not considered suitable for migration as the network latency between the school network and the cloud is too high.
- Teaching: this service is used to host student projects for various courses that require server-side technologies such as MySQL or Apache. This service is only used during term time, which is 24 weeks per year.

Collectively, the above services are currently deployed on 9 application servers and 3 storage servers. The remaining school services are unsuitable for migration as they either control the school network (e.g. the DNS server), or they need low network latencies that make them unsuitable for access over the internet (e.g. day-to-day network storage). A few services require access to hardware or network infrastructure; therefore, they are also unsuitable for migration (e.g. the network monitoring service).

## 4.2 Cost Modeling

The Cost Modeling tool was used as part of the school's cloud migration feasibility analysis. Different models were created to capture the three options that the school is considering (purchasing physical servers, leasing equivalent resources from the cloud, using the elasticity of the cloud). The Cost Modeling tool extends UML deployment diagrams with a UML profile that enables users to model system deployment in the cloud using the following notations:
- Virtual Machine: has an operating system, and either a server type (e.g. AWS.OnDemand.Standard.Small) or server specifications (e.g. CPU clock rate and RAM).
- Virtual Storage: represents persistent storage and can have a type (e.g. AWS.EBS or AWS.S3) in addition to a size (e.g. 100GB) and the number of input and output requests that are expected per month.
- Application: represents software applications that are deployed on virtual machines.
- Data: represents application data that is deployed on virtual storage.
- Database: represents hosted databases such as the Amazon Relational Database Service or Microsoft's SQL Azure.
- Remote Node: represents a node outside the cloud such as an in-house server.
- Communication Path: represents a communication link between any pair of nodes.

- Deployment: represents the deployment of an application onto a virtual machine or the deployment of data onto virtual storage.

Figure 2 shows the deployment model that was created to represent the school's services that are being considered for migration. The model represents the school's network as a remote node that communicates with a monitoring server on the cloud. The services mentioned in Section 4.1 are modeled as applications and data, which are deployed on virtual machines and virtual storage. The model was created using the tool's cloud deployment UML profile, which was installed in the Eclipse IDE. It should be noted that the interdependencies between the applications and other services have been deliberately left out of the diagram to keep things simple and understandable. In practice, these interdependencies do not need to be taken into account during cost modeling as they do not affect costs. The main connections that affect costs are communication paths, which have been included in the model.

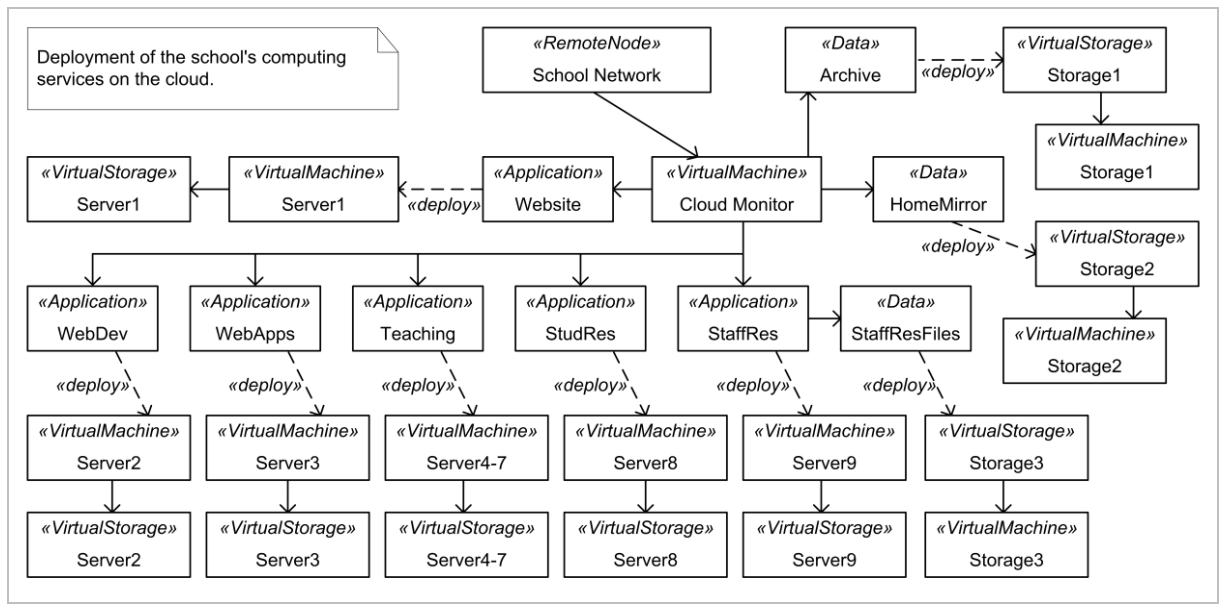

Figure 2: Overview of the school's systems being considered for migration

Once the user has created the model, they can select the cloud provider they wish to use for each of their virtual machines, virtual storage devices or databases. The school is currently considering using Amazon Web Services; however, the tool also supports Microsoft Azure, FlexiScale, Rackspace, and GoGrid (other providers can easily be added). The various infrastructure prices of the cloud providers could have automatically been added to the tool if the providers had created web services that provided the prices; however, they do not currently provide such web services and the prices had to be manually entered into the tool from the providers' websites. The tool has price details for the following resources:
1. Running Hours: the cost of running a virtual machine for one hour.
2. Storage: the cost of storing 1GB of data for one month.
3. Input Requests: the cost of an input request into storage. For some types of storage such as AWS.S3, the cost of a single PUT operation; for other types such as AWS.EBS, this is the cost of a single disk write request (the Unix `iostat` command is useful when obtaining estimates of this figure).
4. Output Requests: the cost of an output request from storage. Depending on which type of storage is being used, this can be the cost of a single GET operation or the cost of a single disk read request.

5. Data In: the cost of transferring 1GB of data into the cloud.
6. Data Out: the cost of transferring 1GB of data from the cloud to another location.

The above resources account for the usage of virtual instances, storage, databases, and data transfer, which are the basic components of any system being deployed in the cloud. There can be other costs associated with running a system in the cloud, e.g. the cost of a static IP address; however, these costs are usually insignificant. Some systems use special services in the cloud, for example Amazon's Cloud Front service that provides fast multimedia content delivery over the web. The Cost Modeling tool does not currently support such services as they are specific to each cloud provider and do not generalize well across various providers. However, support for these special services could be added to the tool.

The key benefit of using the cloud is elasticity and we have developed a simple notation to allow elasticity requirements to be expressed. The tool enables users to define a baseline usage for each resource. Variations to this baseline can be defined using 'usage patterns' that are expressed in natural language. Each pattern could either be temporary or permanent. A temporary pattern is only applied during the month(s) that it is applicable, and can be used to define temporary peaks or drops in usage. In contrast, the resource usage that is changed by a permanent pattern is persistent. Therefore, permanent patterns can be used to define patterns of linear or exponential resource increases or decreases. A pattern is defined as follows:

```
[temp/perm]: every [months] on [days] [variation][number]
```
Where *months*, *days*, *variation* and *number* can be:

| Months  | Days     | Variation | Number           |
|---------|----------|-----------|------------------|
| month   | [empty]  | +         | Float or integer |
| jan-dec | everyday | -         |                  |
|         | weekdays | *         |                  |
|         | weekends | /         |                  |
|         | 01-30    | ^         |                  |
|         | mon-sun  |           |                  |

For example, the following patterns describe a scenario where initially 100GB of storage is required; every month this is increased by 10GB; during weekends between June and August, the required storage is halved; and every December between 25$^{th}$ and the 30$^{th}$, it is doubled.
```
Baseline: 100
Patterns: perm: every month +10, temp: every jun-aug on
weekends /2, temp: every dec on 25-30 *2
```

After a cloud deployment model has been created and the usage patterns have been defined, the user has to set a start and end date for the cost simulations to be performed. Once the simulation starts, the tool converts the graphical deployment model into an XML file that is then used to create a directed cyclic graph representing the model. The usage patterns of each node and edge in the graph are processed for each month between the start and end date of the simulation. The total resource usage of each node is then multiplied by the per-unit cost of that resource, depending on which cloud provider is specified by the user. The per-unit price is retrieved from an XML file that stores the prices from the cloud providers. This file currently contains over 600 prices from various cloud providers.

Finally, the tool generates a detailed cost report showing how the cost of the system changes over time. Figure 3 shows a screenshot of an example report (the screenshot is provided to illustrate the tools UI and need not be read in detail). The school's cost details are analyzed in the next section. The report is a webpage with embedded graphs and tables as well as a zoomable version of the model, which can be very useful when dealing with systems that have a large number of nodes. The system can also export the full costing details as a CSV table for further analysis in Microsoft Excel.

The model can be divided into different groups, and the report provides a detailed breakdown of the costs of each group. A group can represent a department, an organisation or an entire system. This enables architects to evaluate different deployment options of a system and see which is the cheapest. For example, system architects can investigate the costs of duplicating parts of the system on a different cloud for increased availability.

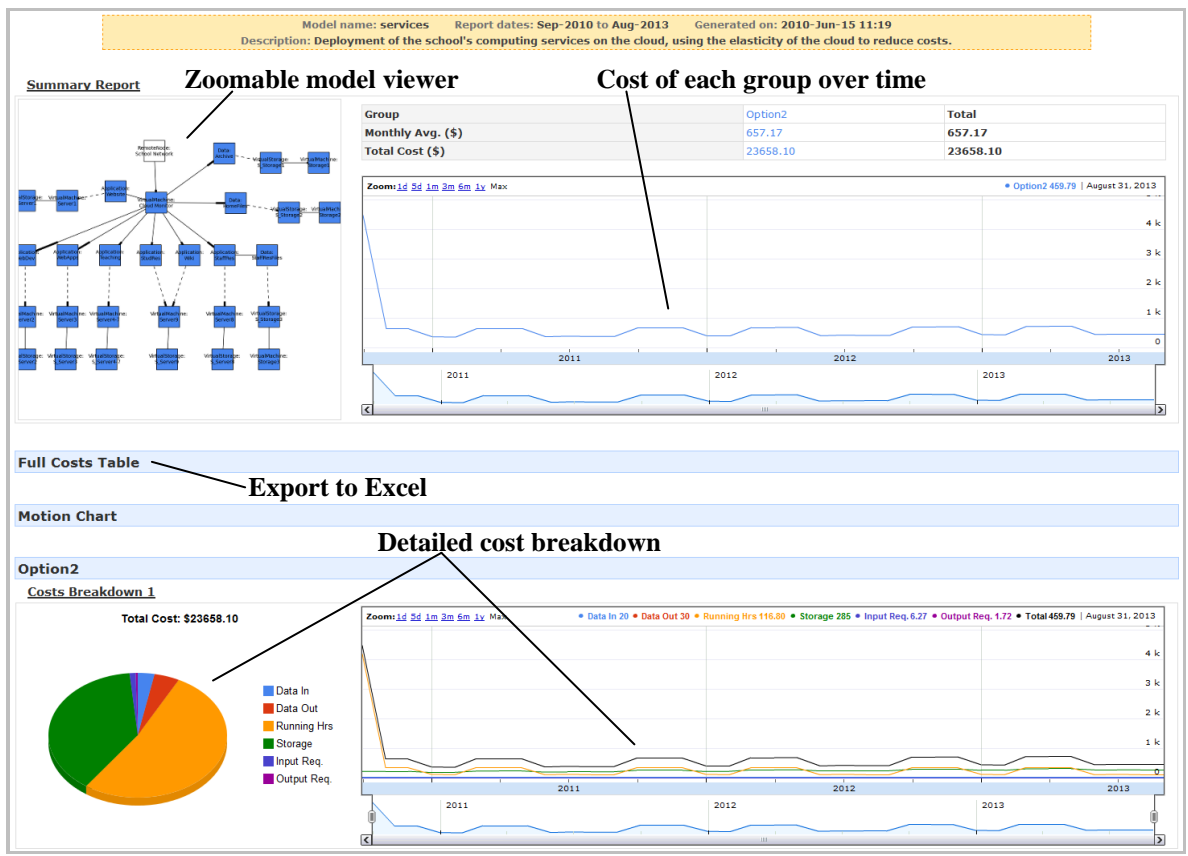

Figure 3: Screenshot of an example cost report showing how the cost of a system could vary over time.

## 4.3 Results

The Cost Modeling tool was used to compare the costs of the school's options over a 6 year period starting from September 2010 (i.e. the start of the next academic year). The school is considering using Amazon Web Services as their cloud provider; therefore, the cost estimates presented in this section are based on AWS's prices; other providers' prices are similar.

**Option 1 - Purchasing physical servers:** 9 application servers and 3 storage servers would be required to replace the existing servers. A mid-range application server costs around $1550 in the UK (e.g. a Dell PowerEdge R410 with an Intel Xeon 2GHz quad-core CPU, 2GB RAM and two 250GB hard drives configured in RAID1 to give 250GB usable storage). A mid-range storage server costs around $2500 (e.g. a Dell PowerEdge R510 with an Intel Xeon 2GHz quad-core CPU, 4GB RAM and five 250GB hard drives configured in RAID5 to give 1TB usable storage, with an extra disk as a hot spare). Electricity costs would be $106 per year for each application server and $155 per year for each storage server (based on energy usage estimates from Dell[6], assuming a 10% CPU load and a cost of $0.1 per kWh, which is what the school pays). Cooling and network infrastructure costs do not need to be considered as the school already has these facilities in its machine room for the existing servers. The costs of purchasing physical servers were calculated using a 3-year upgrade cycle where the school would pay the same upfront capital to upgrade to new servers in year 4. This is a reasonable upgrade cycle for the purposes of cost comparison as any server failures during this time are covered by Dell's basic 3-year guarantee.

**Option 2 - Leasing equivalent resources from the cloud:** Leasing the equivalent amount of resources in option 1 from the cloud would require 12 HighCPU.Medium instances from Amazon EC2's European region (using 'reserved 3-year' instances to reduce costs). The reserved instances option would have to be renewed in year 4 to keep the instance costs low. Similarly to option 1, each application server would have a 250GB EBS volume, and each storage server would have a 1TB EBS volume. The number of I/O operations were measured on the existing servers and these values were input into the Cost Modeling tool. In addition, it was estimated that 200GB of data would be transferred into the cloud each month, and 200GB would be transferred out each month.

**Option 3 - Using the elasticity of the cloud:** The resource usage of the existing servers was reviewed and the cost model that was created for option 2 was modified to include the school's actual resource usage. This involved defining patterns to switch-off instances when they were not in use. For example, the baseline number of instances for the teaching service was set to 0, and the usage pattern was set to `[temp: every sep-nov +4, temp: every feb-apr +4]` to show that 4 servers would be required during term time. Three of the school's services did not require the HighCPU.Medium type of instance as they had a small usage; therefore they were deployed on Standard.Small instances. In addition, the storage servers were replaced by using Amazon's S3 service, and usage patterns were defined to show how the school's storage demands increase over time. For example, the baseline storage of the archive service was set to 560GB, and the usage pattern was set to `[perm: every month +15]` to show that 15GB of extra storage would be required every month.

Amazon has previously changed their pricing scheme for some their services, for example in November 2009 they lowered the price of all on-demand instances by 15%[7]. Therefore, it is useful for decision makers to consider the cost of their systems if cloud providers change their pricing scheme in the future. The school was interested to see how the cost of their system would change, if in 2 years time, Amazon:
1. Increases instance-hour and storage prices by 15% due to rising energy costs.
2. Decreases instance-hour and storage prices by 15% due to Moore's Law and more powerful hardware coupled with increasing competition from other cloud providers.

Figure 4 shows how much the school would be paying for each option over the 6 year period that is being investigated. At the start (i.e. year 0), it would either cost $22,800 to buy physical servers (includes electricity usage for first year) or $23,300 to lease equivalent resources in the cloud. However, if the system is modified to use the elasticity of the cloud, then the starting cost would be $9,900. Figure 4 shows how the costs vary over the remaining years, for example in year 1, the elastic option would cost $6,700 compared to $1,400 for the electricity usage of the buy option.

---

[6] http://solutions.dell.com/DellStarOnline/DCCP.aspx
[7] http://aws.amazon.com/about-aws/whats-new/2009/10/27/announcing-lower-amazon-ec2-instance-pricing/

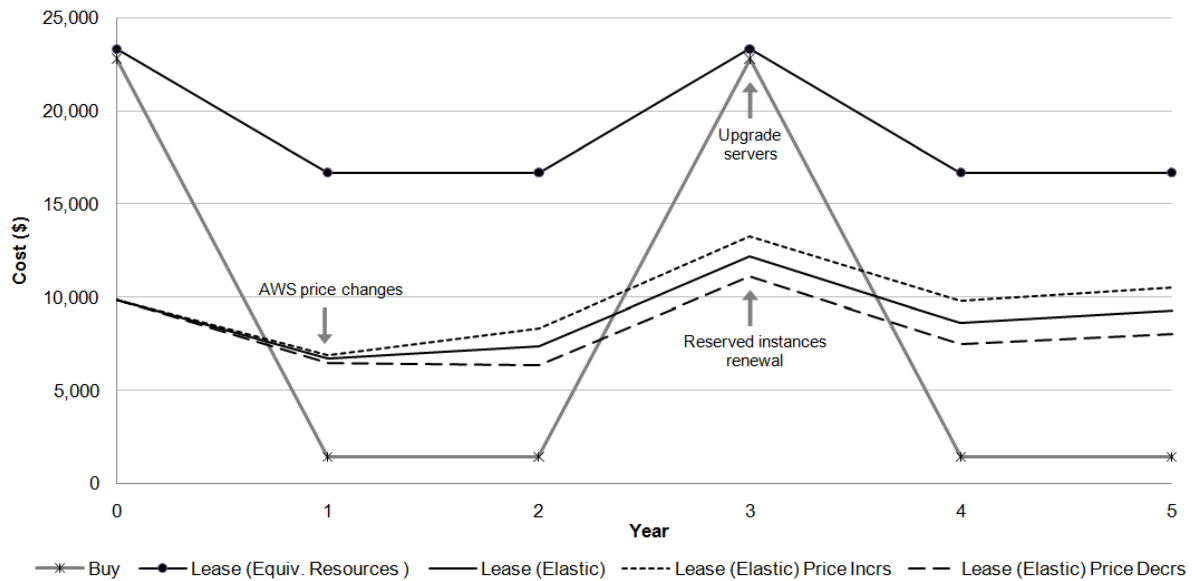

Figure 4: The yearly cost of different options that the school could take

To compare the school's options financially, their *net present values* (NPV) have to be calculated over the 6 years. NPV is often used by organisations to compare the overall value of different investment options by taking into account their incoming and outgoing cash flows [35]. Since the school does not make explicit profits from its computing services, the incoming cash flow can be ignored. NPV calculations take into account the *cost of capital*, which is the return rate that capital could earn in an alternative investment option [35]. For example, the school could put the upfront capital into a bank savings account and earn interest if they choose a lease option. Assuming a 5% return rate, each cost, *C*, at year *Y* in Figure 4 has to be set to: $C = C / (1 + 0.05)^Y$. These costs then have to be summed to give the NPV of each scenario, which is shown in Figure 5. The percentage differences between the buy option and all other options are also shown in the figure.

It should be noted that a higher return rate favors the cloud option as future costs become more rewarding than upfront costs. Surprisingly, it can be seen that the elastic option is slightly more expensive than buying physical servers for the school. Leasing equivalent resources from the cloud and leaving them running 24x7 makes no financial sense as it costs more than twice the buy option. However, if Amazon reduces prices by 15% in 2 years time, then the elastic option becomes the cheapest option.

We did not explore the possible option of buying fewer physical servers and using virtualization to run several servers on one machine. This would certainly have reduced the overall costs of purchase but would incur additional local setup costs. Nor did we take account of any changes to staffing required – in practice, we do not think that there would be any significant reduction in support costs.

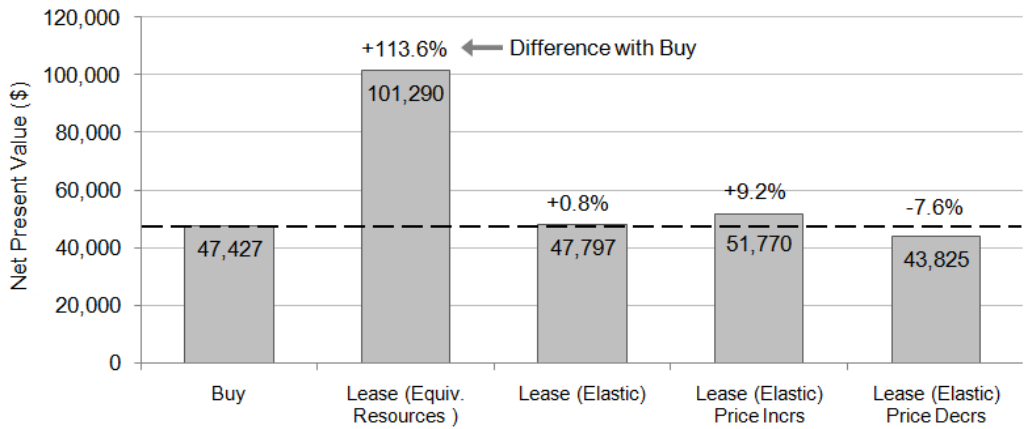

Figure 5: The net present value of different options that the school could take

## 4.4 Discussion

This case study demonstrates that: the Cost Modeling tool addresses the challenges of cost analysis (as defined in Section 2); that the output of Cost Modeling informs cloud adoption decisions by providing important information to decision makers; and that the cost effectiveness of public clouds is situation dependent rather than universally less expensive than traditional forms of IT provisioning. Finally, we discuss the implications of these results within the context of our previous findings that suggest that practitioners should be careful not to ignore organizational change issues and rely solely upon financial data in making decisions.

The case study demonstrates that the Cost Modeling tool addresses the challenges of determining the cost of operating a typical subset of enterprise IT infrastructure in the cloud:
- It models the actual resources used by a system, e.g. storage, bandwidth, CPU usage.
- It models the pricing schemes of multiple cloud providers.
- It scales to model the IT infrastructure of a typical enterprise environment.
- It is easy to use as it requires no coding as the models can be defined graphically using existing UML modeling applications such as the Eclipse IDE.
- It is platform independent having been implemented using Eclipse, Python and a combination of open source libraries.

The results demonstrate that the output of the Cost Modeling tool helps to inform migration decisions. The output recommended that the school should buy physical servers if they have the upfront capital. If not, then they should lease resources from a cloud provider but re-architect their system to use the cloud's elasticity, otherwise the costs would be higher than buying physical servers. We found that due to situation specific factors, the results of the tool needed to be supplemented. For example, the *opportunity cost* of the buy option's upfront capital should also be considered. That is the benefit that the school would have received if they had used that capital to take an alternative action [35]. For instance, as the elastic option needs 60% less capital upfront, the remaining capital could be used for other investments such as improving facilities or increasing the publicity budget to recruit more students.

Another factor that should also be considered is the cost of infrastructure support and maintenance. Servers that are used for business critical applications often require expensive support and maintenance contracts with hardware suppliers that guarantee response times to support calls. Cloud providers are beginning to address this demand as well, for example Amazon has a premium support package[8] that guarantees a one-hour response time for urgent issues. The individual services offered by each cloud provider and their service level agreements (SLA) should also be considered, in addition to the compensation that is provided if they fail to meet their SLA.

An important limitation to any cloud cost estimation approach (including Cost Modeling) is the need to have fairly accurate estimates of resource usage, as the estimated costs are sensitive to inaccuracies. For example, in this case study we identified that if a summer school is being run and it requires the use of 4 servers, then these would have to be leased from the cloud. In contrast, if the school already has 4 teaching servers that are not utilized during the summer, then those could be used. With traditional infrastructure provisioning enterprises do not have to worry too much about usage patterns when they buy servers as they are often underutilized [36] and can accommodate temporary peaks. However, public cloud infrastructure could be perfect for situations where usage patterns are unknown, and resource needs cannot be assumed to be met by underutilized servers. For example, if the school is considering introducing distance learning courses but they do not know the level of demand for such courses, then using the cloud makes financial sense as buying servers would be too risky. If eventually there is enough demand that is fairly constant and continuous in nature, then it could be case that the school could actually save money by migrating the courses into in-house servers.

The results of this case study can be understood within the context of our existing work as re-enforcing our arguments that decision makers should not rely solely upon financial data when making decisions pertaining to the adoption of cloud. Our previous case study that investigated the migration of system infrastructure to Amazon EC2 in an oil and gas services company showed that the system infrastructure would have cost around 37% less over 5 years on EC2 compared to the in-house data centre [4]. In contrast, the results of this case study showed that despite popular beliefs of cost savings in the cloud, there is not much difference between the costs of buying physical servers and the costs of deploying some of the school's IT systems on the cloud. The difference between the two case studies is that the system mentioned in [4] was a green-field development project; therefore new network infrastructure had to be purchased. It could well be that the cloud is a cheaper option for an enterprise that needs more than say 30 servers, due to the extra costs of racks, cooling, and network infrastructure that would be required for physical servers. The difference between these two case studies highlights the importance of, and the need for, the Cost Modeling tool to enable decision makers to investigate the costs of deploying their specific systems on the cloud.

However, it should be noted that despite the favorable cost analysis in our previous case study (with the oil and gas services company) it was decided not to migrate to the cloud due to benefits and risks relating to organizational change. These benefits can also be observed in this case study. The system administrators would be freed from maintaining hardware and can focus on supporting applications. The load on the school's internal network could potentially be reduced as requests would be sent to the cloud.

However, there are also barriers to using the cloud, mainly the migration of data and applications, which requires the system to be re-designed to use elasticity. System administrators will require some training for this but it should not require too much effort as only the management of the infrastructure is affected (e.g. Amazon's APIs need to be used). Using elasticity for the school's systems should be fairly straightforward as it will involve switching-off virtual machines that are not in use, and using Amazon S3 is inherently elastic as storage does not need to be provisioned beforehand. In contrast, using elasticity can be challenging and expensive to achieve for interconnected enterprise systems that rely on other systems being available all the time, or for systems that use relational databases that cannot be easily scaled out. There are also the usual security and privacy issues that are often raised [21], but such issues are not significant for this case study as data would be encrypted before being transferred to the cloud for storage. Other data, such as teaching material are already available on the web and therefore the implications of storing them in the cloud would be no different.

---

[8] http://aws.amazon.com/premiumsupport/

# 5    Conclusion

Enterprises are currently at the start of a transition period during which they face many challenges with respect to cloud adoption in the enterprise. This paper argued that understanding the organizational benefits and drawbacks is far from straightforward because: the suitability of the cloud for many classes of systems is unknown or an open-research challenge; cost calculations are complicated due to the number of variables comprising inputs to the utility billing model of cloud computing; the adoption of cloud computing results in a considerable amount of organizational change that will affect peoples' work in significant ways; corporate governance issues regarding the use of cloud computing are not well understood.

Subsequently we demonstrated that there are currently no mature techniques or toolkits available to address the identified challenges, and thus support decision making during the adoption of cloud computing in the enterprise. We argue that our Cloud Adoption Toolkit offers a promising starting point for cloud migration decision making. The Cloud Adoption Toolkit includes Technology Suitability Analysis, Energy Consumption Analysis, Stakeholder Impact Analysis, Responsibility Modeling and Cost Modeling.

We then demonstrated the value of the toolkit using a case study that focused on Cost Modeling (the most mature component of the toolkit), and showed that: the Cost Modeling tool addresses the challenges of cost analysis; that the cost saving estimates that are often cited by cloud providers cannot be generalized across all IT systems as they very much depend on the specific resource usage and the deployment options being used by a system. Based on the results of the Cost Modeling tool, it was recommended that the organization involved in the case study should buy physical servers if they have the upfront capital. If not, then they should lease resources from a cloud provider but re-architect their system to use the cloud's elasticity, otherwise the costs would be higher than buying physical servers. These findings were then placed in the context of our earlier findings [4] that demonstrated that cost savings can be established in specific classes of systems, and that a purely financial analysis of a cloud adoption is inadequate as it fails to address important issues of organizational change. Our findings are limited by the usual drawbacks of case studies as such it is difficult to generalize from individual cases. However, as discussed, these results corroborate our previous findings, and suggest a more general trend.

As part of future work we intend to use the toolkit's Stakeholder Impact Analysis technique to investigate the impact of cloud adoption by the organization that was involved in the case study. Responsibility Modeling was not required in the case study as only one organization and one cloud provider would be involved in maintaining and operating the system. Energy consumption analysis was also not required in the case study because the organization was only considering the use of public clouds. We are currently considering exposing the functionality of the Cost Modeling tool as a web application to open the tool to the practitioner community. We are also considering creating UML profiles for widely used UML modeling applications such as Rational Rose and exposing the functionality of the Cost Modeling tool as a web service to enable users to send their cloud deployment models from their UML applications and receive detailed cost reports.

## Acknowledgements

We thank the Scottish Informatics and Computer Science Alliance (SICSA) and the EPSRC (grant numbers EP/H042644/1 and EP/F001096/1) for funding this work. We also thank our colleagues at the UK's Large-Scale Complex IT Systems Initiative for their comments.